# Business and consumer uncertainty in the face of the pandemic: A sector analysis in European countries

Oscar Claveria[*]

AQR-IREA, Department of Econometrics, Statistics and Applied Economics, University of Barcelona

**Abstract.** This paper examines the evolution of business and consumer uncertainty amid the coronavirus pandemic in 32 European countries and the European Union (EU). Since uncertainty is not directly observable, we approximate it using the geometric discrepancy indicator of Claveria et al. (2019). This approach allows us quantifying the proportion of disagreement in business and consumer expectations of 32 countries. We have used information from all monthly forward-looking questions contained in *Joint Harmonised Programme of Business and Consumer Surveys* conducted by the European Commission (EC): the industry survey, the service survey, the retail trade survey, the building survey and the consumer survey. First, we have calculated a discrepancy indicator for each of the 17 survey questions analysed, which allows us to approximate the proportion of uncertainty about different aspects of economic activity, both form the demand and the supply sides of the economy. We then use these indicators to calculate disagreement indices at the sector level. We graphic the evolution of the degree of uncertainty in the main economic sectors of the analysed economies up to June 2020. We observe marked differences, both across variables, sectors and countries since the inception of the COVID-19 crisis. Finally, by adding the sectoral indicators, an indicator of business uncertainty is calculated and compared with that of consumers. Again, we find substantial differences in the evolution of uncertainty between managers and consumers. This analysis seeks to offer a global overview of the degree of economic uncertainty in the midst of the coronavirus crisis at the sectoral level.

[*] **Corresponding Author:**
Oscar Claveria, University of Barcelona, 08034 Barcelona, Spain. Email: oclaveria@ub.edu

# 1. Introduction

The analysis of economic uncertainty gains renewed interest since the advent of the coronavirus pandemic and the subsequent economic disruption caused by the lockdown. There is ample evidence that uncertainty shocks have an effect on real activity (Baker et al. 2016; Bloom 2009; Henzel and Rengel 2017). Since economic uncertainty is not directly observable, several strategies have been designed to proxy it by: a) using the realized volatility in equity markets (Bekaert et al. 2013; Caggiano et al. 2014), b) estimating econometric unpredictability –understood as the conditional volatility of the unforecastable components of a broad set of economic variables (Jurado et al. 2015; Meinen and Roehe 2017)–, and c) computing survey-derived measures of expectations dispersion (Clements and Galvão 2017; Krüger and Nolte 2016). The ex-ante nature of this latter approach has generated a growing current in the literature based on this type of metrics (Dovern 2015; Mankiw et al. 2004; Meinen and Roehe 2017).

Disagreement measures based on survey expectations make use of prospective information, as agents are asked about the expected future evolution of a wide range of variables. While most studies rely on quantitative macroeconomic expectations made by professional forecasters (Lahiri and Sheng 2010; Oinonen and Paloviita 2017), an alternative source of survey expectations are business and consumer tendency surveys (Claveria, 2020, 2021; Binding and Dibiasi 2017; Das et al. 2019; Shainoz and Cosar 2020). See Grimme et al. (2014) for the evaluation of a combination of different types of measures.

The European Commission (EC) conducts monthly business and consumer tendency surveys in which respondents are asked whether they expect a set of economic variables to rise, fall or remain unchanged. We use all the forward-looking information coming from these surveys to proxy economic uncertainty in 32 European countries and the European Union (EU). To this end, we use Claveria et al.'s (2019) geometric indicator of discrepancy to compute the proportion of disagreement among firms and households.

Given that survey expectations: (a) are based on the knowledge of respondents operating in the market, (b) provide detailed information about a wide range of economic variables, and (c) are available ahead of the publication of official quantitative data, the proposed approach to measure economic uncertainty allows us to give a quick snapshot of economic uncertainty amid the COVID-19 pandemic in real time.

The main aim of the study is to provide some insight regarding the recent evolution of uncertainty across economic sectors, economic agents and countries, both from the demand and the supply sides of the economy.

The next section describes the data and describes the methodological approach to compute disagreement among agents. The graphical analysis is provided in Section 3.

**2. Data and Methodology**

We use firms' and consumers' qualitative expectations about a wide array of economic variables (see Table 1). Specifically, we use all forward-looking monthly raw data from all business and consumer surveys conducted by the EC. The sample period goes from 2016.M5 to 2020.M2 since we wanted to focus on the evolution of disagreement during the months previous to the coronavirus pandemic. This allowed us to include all the available information from all the surveys in all the 32 economies in which the surveys are now conducted. To our knowledge, this is the first study that includes Montenegro, North Macedonia, Albania, Serbia and Turkey, which were recently added to the survey.

**Table 1.** Survey indicators

| **Industry survey** |
|---|
| *I5* – Production expectations for the months ahead |
| *I6* – Selling price expectations over the next 3 months |
| *I7* – Employment expectations over the next 3 months |
| **Service survey** |
| S3 – Expectation of the demand over the next 3 months |
| *S5* – Expectations of the employment over the next 3 months |
| *S6* – Expectations of the prices over the next 3 months |
| **Retail trade survey** |
| *R3* – Orders expectations over the next 3 months |
| *R4* – Business activity expectations over the next 3 months |
| *R5* – Employment expectations over the next 3 months |
| *R6* – Prices expectations over the next 3 months |
| **Building survey** |
| *B4* – Employment expectations over the next 3 months |
| *B5* – Prices expectations over the next 3 months |
| **Consumer survey** |
| *C2* – Financial situation over next 12 months |
| *C4* – General economic situation over next 12 months |
| *C6* – Price trends over next 12 months |
| *C7* – Unemployment expectations over next 12 months |
| *C9* – Major purchases over next 12 months |

In business surveys, respondents are asked about their expectations regarding firm-specific factors such as production, selling prices and employment and, they are faced with three options: "up", "unchanged" and "down". $P_t$ measures the share of respondents reporting an increase in the variable, $E_t$ no change, and $M_t$ a decrease. The most common way of presenting survey data is the balance, $B_t$, which is computed as the subtraction between the two extreme categories: $B_t = P_t - M_t$.

Consumers, for their part, are asked about objective variables (e.g. how they think the general economic situation in the country will change over the next twelve months) and subjective variables (e.g. major purchases, savings, etc.). Consumers have three additional response categories: two at each end ("a lot better/much higher/sharp increase", and "a lot worse/much lower/sharp decrease"), and a "don't know" option. As a result, $PP_t$ measures the percentage of respondents reporting a sharp increase in the variable, $P_t$ a slight increase, $E_t$ no change, $M_t$ a slight fall, $MM_t$ a sharp fall and, $N_t$ don't know.

The most widespread measures of disagreement among survey respondents use the dispersion of balances as a proxy for uncertainty (Bachmann et al. 2013; Girardi and Reuter 2017; Mokinski et al. 2015). Bachmann et al. (2013) proposed an indicator of disagreement based on the square root of the variance of the balance:

$$DISP_t = \sqrt{P_t + M_t - (P_t - M_t)^2} \qquad (1)$$

See Dibiasi and Iselin (2019) for a comparison of (1) to Theil's disconformity coefficient (Theil 1955) and analysis of firms' direct perception of investment uncertainty. By means of a simulation experiment, Claveria et al. (2019) showed that the omission of neutral responses in (1) resulted in an overestimation of the level of disagreement. As a result, the authors developed a disagreement metric that incorporated the information coming from all the reply options (*N*). Given that the sum of the shares adds to one, the authors computed an *N*-dimensional vector encompassing all shares, and projected it as a point on a simplex of $N-1$ dimensions. For $N=3$, the simplex takes the form of an equilateral triangle, where the point corresponds to a unique convex combination of the three reply options for each period in time. See Claveria (2018) for an extension of the methodology for a larger number of reply options, and Claveria (2019) for an application of the methodology when $N=5$.

Insomuch as all vertices are at the same distance to the centre of the simplex ($O$), the ratio of the distance of a point to the barycentre ($VO$) and the distance from the barycentre to the nearest vertex ($OP$) provides the proportion of agreement among respondents. Consequently, the indicator of discrepancy for a given period in time can be formalised as:

$$D_t = 1 - \left[ \frac{\sqrt{(P_t - 1/3)^2 + (E_t - 1/3)^2 + (M_t - 1/3)^2}}{\sqrt{2/3}} \right] \quad (2)$$

This metric is bounded between zero and one, and conveys a geometric interpretation. The center of the simplex corresponds to the point of maximum disagreement, indicating that the answers are equidistributed among the three response categories. Conversely, each of the $N$ vertexes corresponds to a point of minimum disagreement, where one category draws all the answers and $D_t$ reaches the value of zero.

When comparing the evolution of the geometric measure of disagreement (2) to that of the standard deviation of the balance (1) in several European countries, Claveria (2020) obtained a high positive correlation between both measures of disagreement, and found that the main difference between both measures mainly lied in their average level and dispersion, being *DISP* more volatile and higher in most countries. In this study we apply expression (2) to measure discrepancy in all business and consumer surveys.

## 3. Graphical analysis

In this section we use qualitative survey data from the five independent tendency surveys conducted by the EC – the industry survey (INDU), the service survey (SERV), the retail trade survey (RETA), the construction survey (BUIL), and the consumer survey (CONS) – to compute the proportion of disagreement among respondents. By averaging the information coming from the different variables in each survey, we compute sector indicators of disagreement, which we in turn use to compute a business disagreement indicator that aggregates the information coming from the four sector indicators. We use all these indicators to examine the evolution of uncertainty, both from the demand (Fig.1) and the supply sides of the economy (Fig.2). Finally, in Table 2 we ranked the countries according to their average values of disagreement across the sample. See the Appendix for a detailed descriptive analysis of disagreement for all variables and surveys.

**Fig. 1a.** Evolution of industry, service, retail trade and construction disagreement

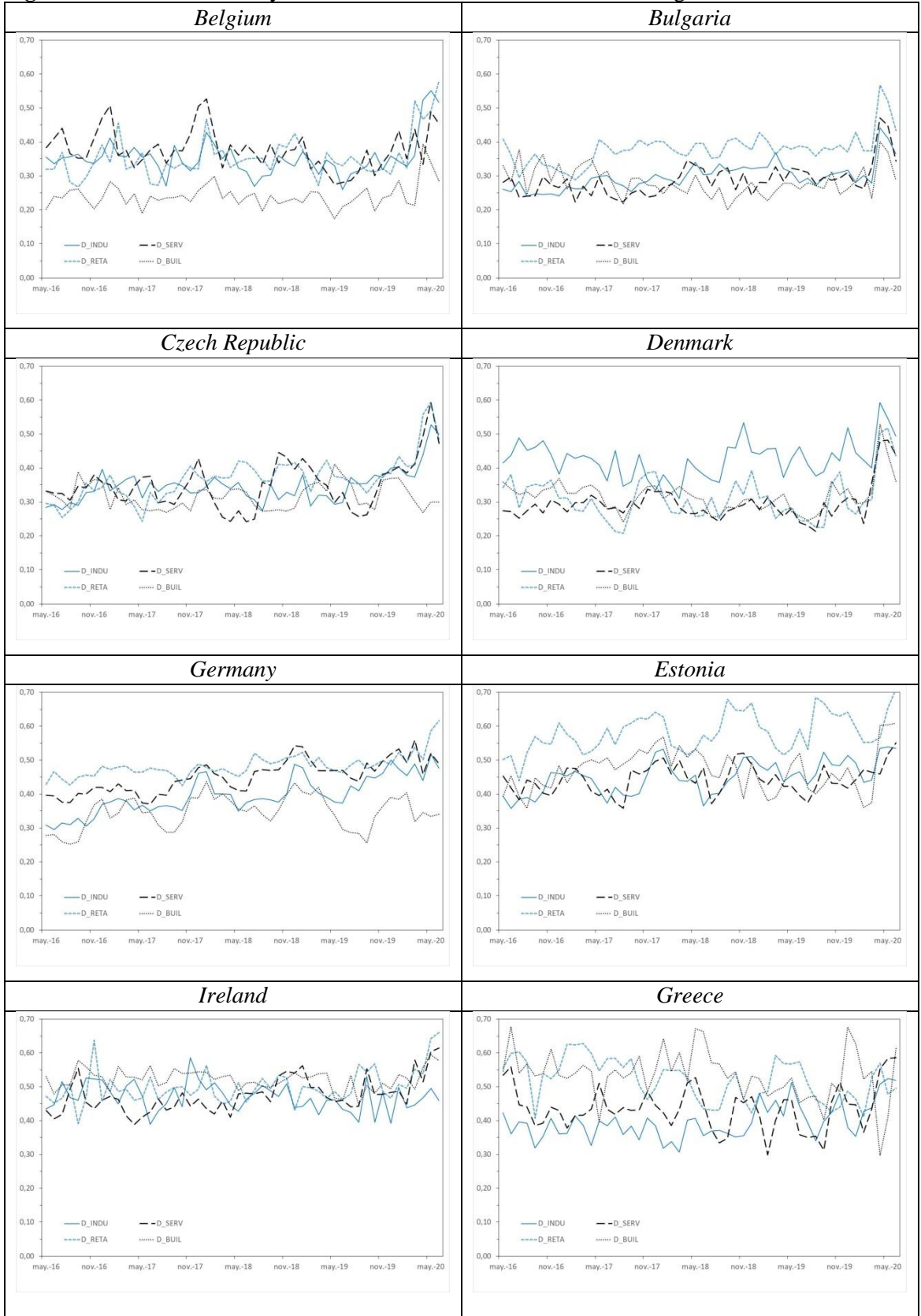

Notes: The solid blue line represents the evolution of industry disagreement, the dashed black line the evolution of service disagreement, the dashed blue line the evolution of retail trade disagreement, and the dotted black line the evolution of construction disagreement.

**Fig. 1b.** Evolution of industry, service, retail trade and construction disagreement

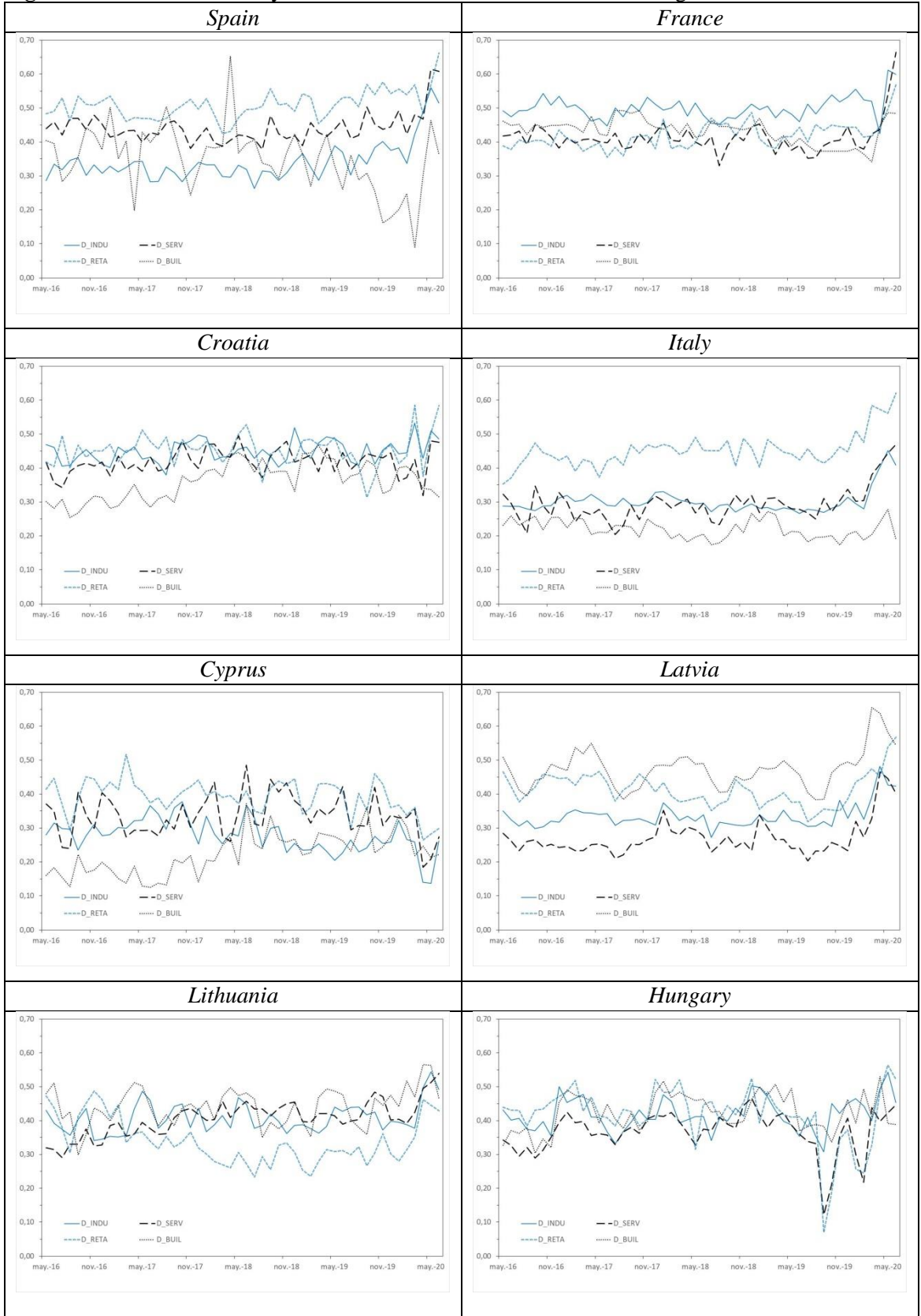

Notes: The solid blue line represents the evolution of industry disagreement, the dashed black line the evolution of service disagreement, the dashed blue line the evolution of retail trade disagreement, and the dotted black line the evolution of construction disagreement.

**Fig. 1c.** Evolution of industry, service, retail trade and construction disagreement

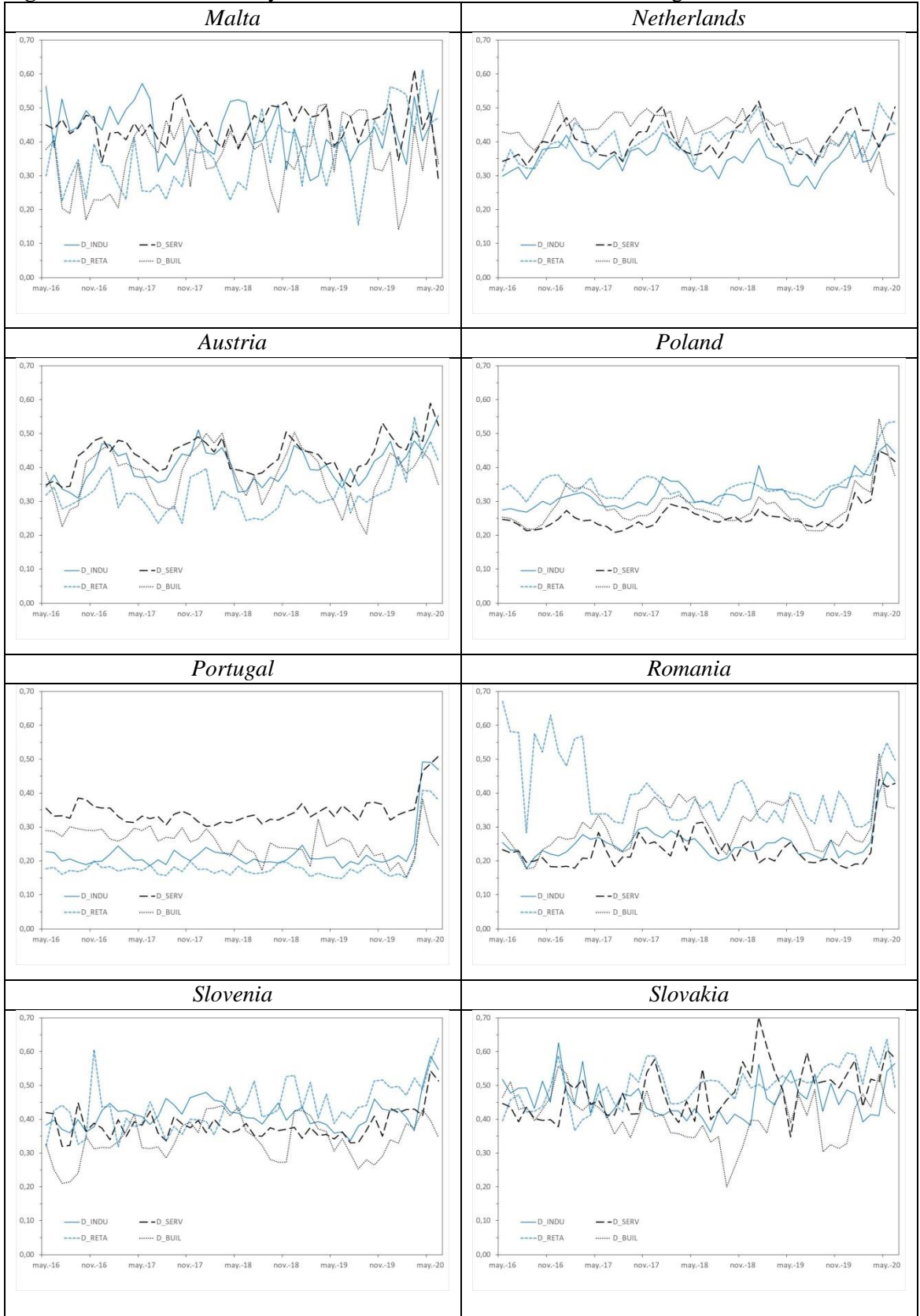

Notes: The solid blue line represents the evolution of industry disagreement, the dashed black line the evolution of service disagreement, the dashed blue line the evolution of retail trade disagreement, and the dotted black line the evolution of construction disagreement.

**Fig. 1d.** Evolution of industry, service, retail trade and building disagreement

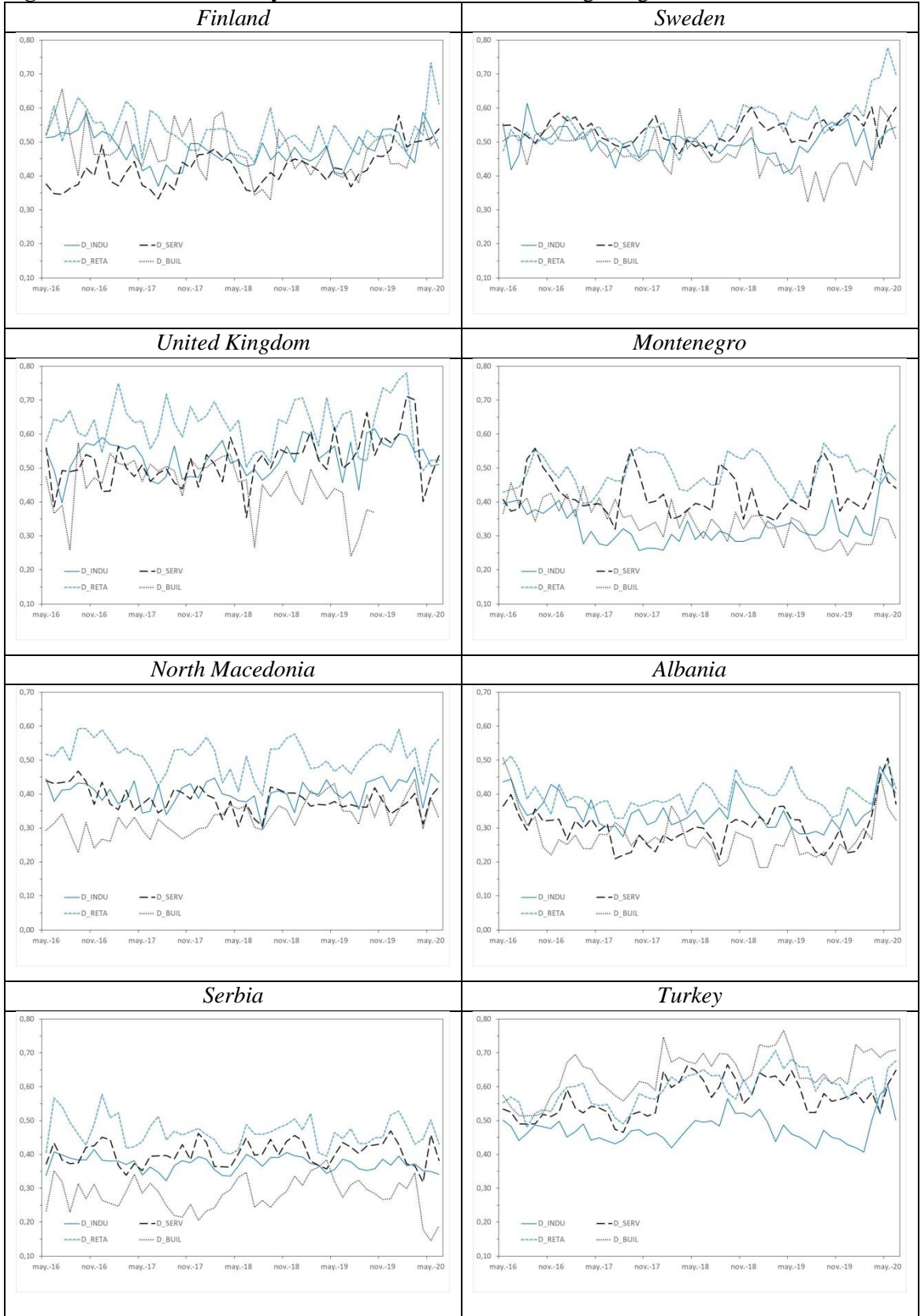

Notes: The solid blue line represents the evolution of industry disagreement, the dashed black line the evolution of service disagreement, the dashed blue line the evolution of retail trade disagreement, and the dotted black line the evolution of construction disagreement.

**Fig. 1e.** Evolution of industry, service, retail trade and construction disagreement

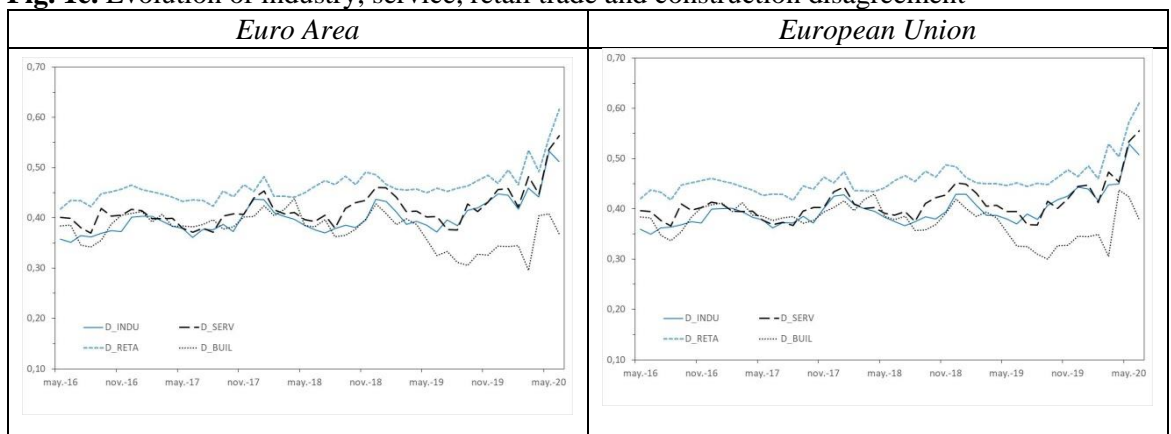

Notes: The solid blue line represents the evolution of industry disagreement, the dashed black line the evolution of service disagreement, the dashed blue line the evolution of retail trade disagreement, and the dotted black line the evolution of construction disagreement.

In Fig.1 we observe that the evolution of disagreement varies both across sectors and countries:

- If we focus on the last months of 2020, when the effects of the first wave of the coronavirus crisis where already palpable, we find different patterns regarding the evolution across sectors.

- In most cases, disagreement in the industry sector (BUIL) starts to decrease in April or May 2020, and in the construction sector even before that. In contrast, disagreement in the service sector and the retail trade sector continues to rise (France, Italy and Estonia).

- In Belgium, Germany, Greece, Spain, Croatia, Latvia, Poland or Slovenia retail trade (RETA) disagreement shows an increasing trend as of June 2020. In Finland, Sweden, Lithuania, the Netherlands, Hungary, Portugal, Montenegro, and North Macedonia it is service disagreement that maintains its growing trend.

- In other countries like Bulgaria, Czechia, Romania, Albania and Denmark, disagreement in all sectors co-evolves, decreasing after April or May 2020, as opposed to Cyprus and the United Kingdom where disagreement in all sectors rises.

- In Austria and Slovakia industry disagreement does not decrease and shows an increasing trend in June 2020. Cyprus, Greece and Turkey are the only countries in which disagreement in the building sector keeps rising in June 2020.

In Fig.2 we compared the evolution of business disagreement vs. consumer disagreement in each economy.

**Fig. 2a.** Evolution of business disagreement vs. consumer disagreement

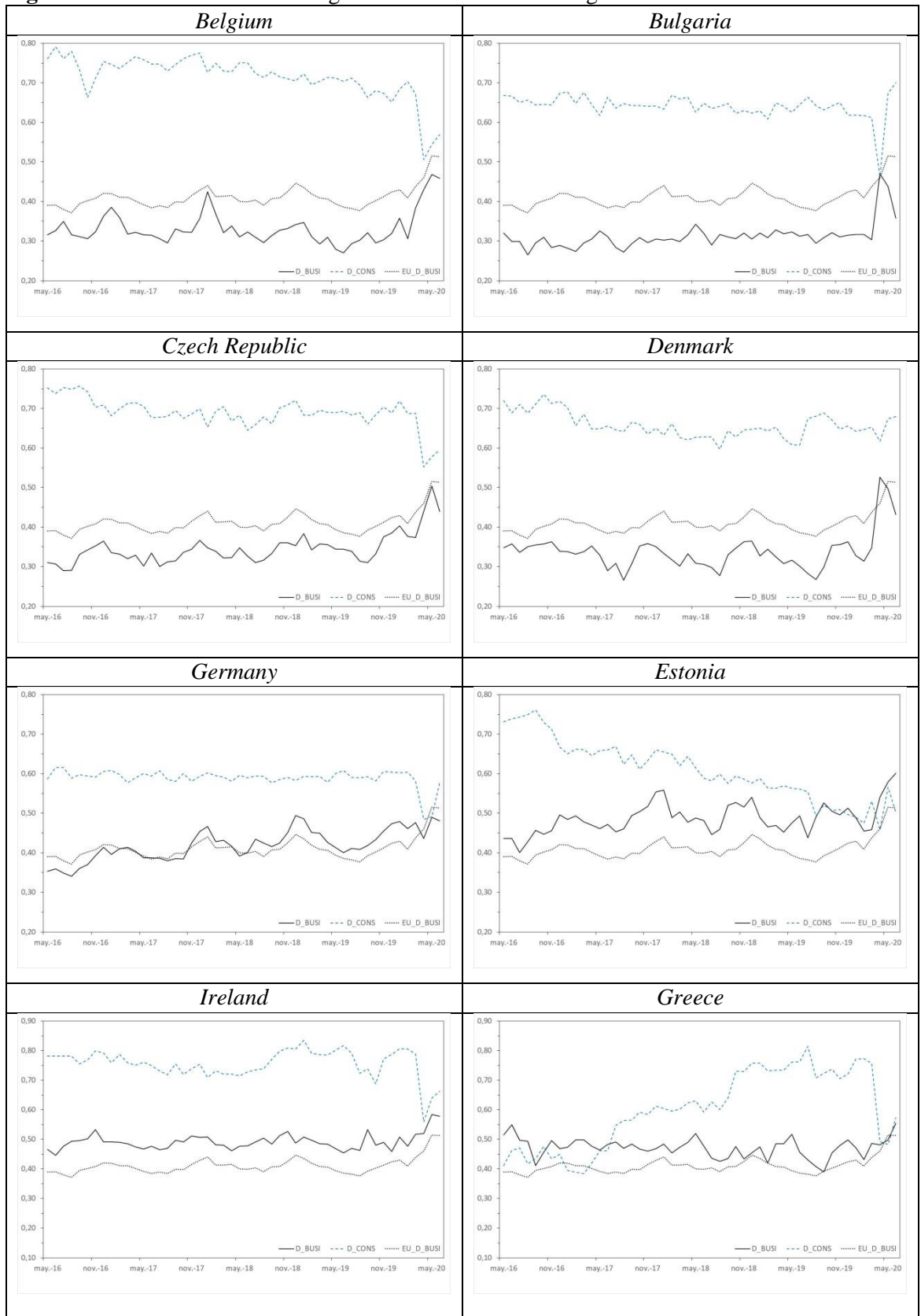

Notes: The solid black line represents the evolution of business disagreement in each country –aggregate disagreement for industry, service, retail trade and construction–, the dashed blue line the evolution of consumer disagreement in each country, and the dotted black line the evolution of aggregate business disagreement in the EU.

**Fig. 2b.** Evolution of business disagreement vs. consumer disagreement

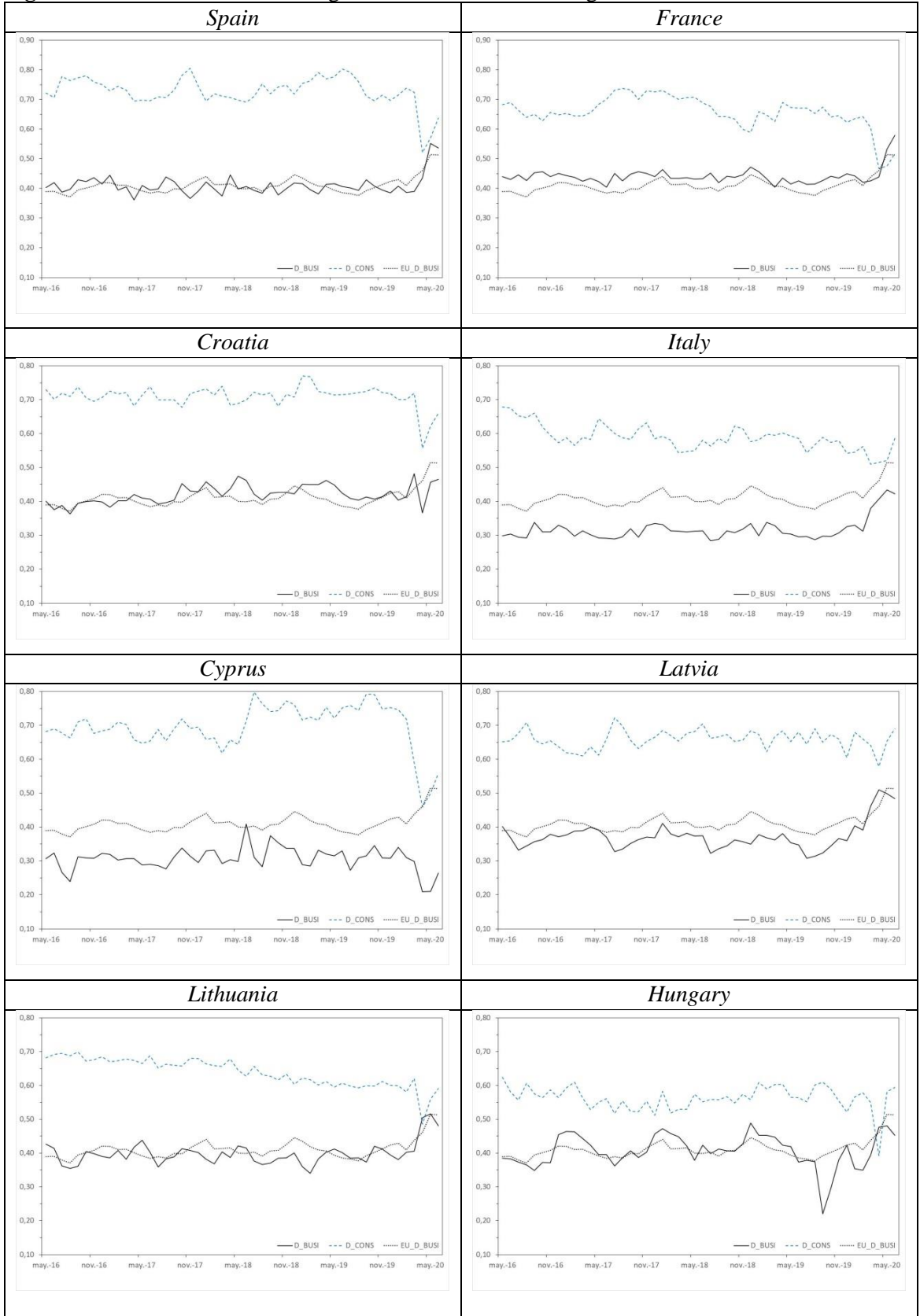

Notes: The solid black line represents the evolution of business disagreement in each country –aggregate disagreement for industry, service, retail trade and construction–, the dashed blue line the evolution of consumer disagreement in each country, and the dotted black line the evolution of aggregate business disagreement in the EU.

**Fig. 2c.** Evolution of business disagreement vs. consumer disagreement

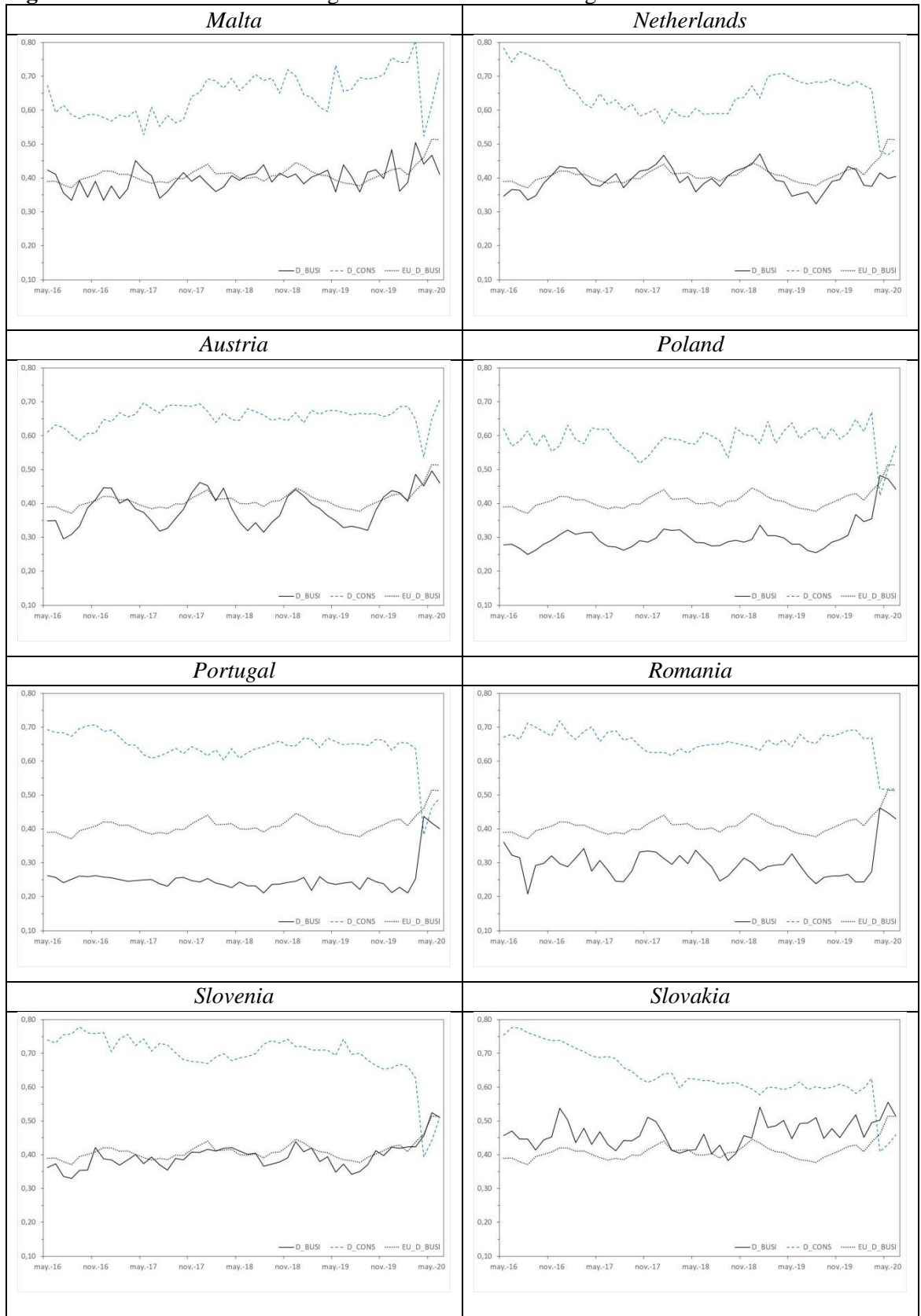

Notes: The solid black line represents the evolution of business disagreement in each country –aggregate disagreement for industry, service, retail trade and construction–, the dashed blue line the evolution of consumer disagreement in each country, and the dotted black line the evolution of aggregate business disagreement in the EU.

**Fig. 2d.** Evolution of business disagreement vs. consumer disagreement

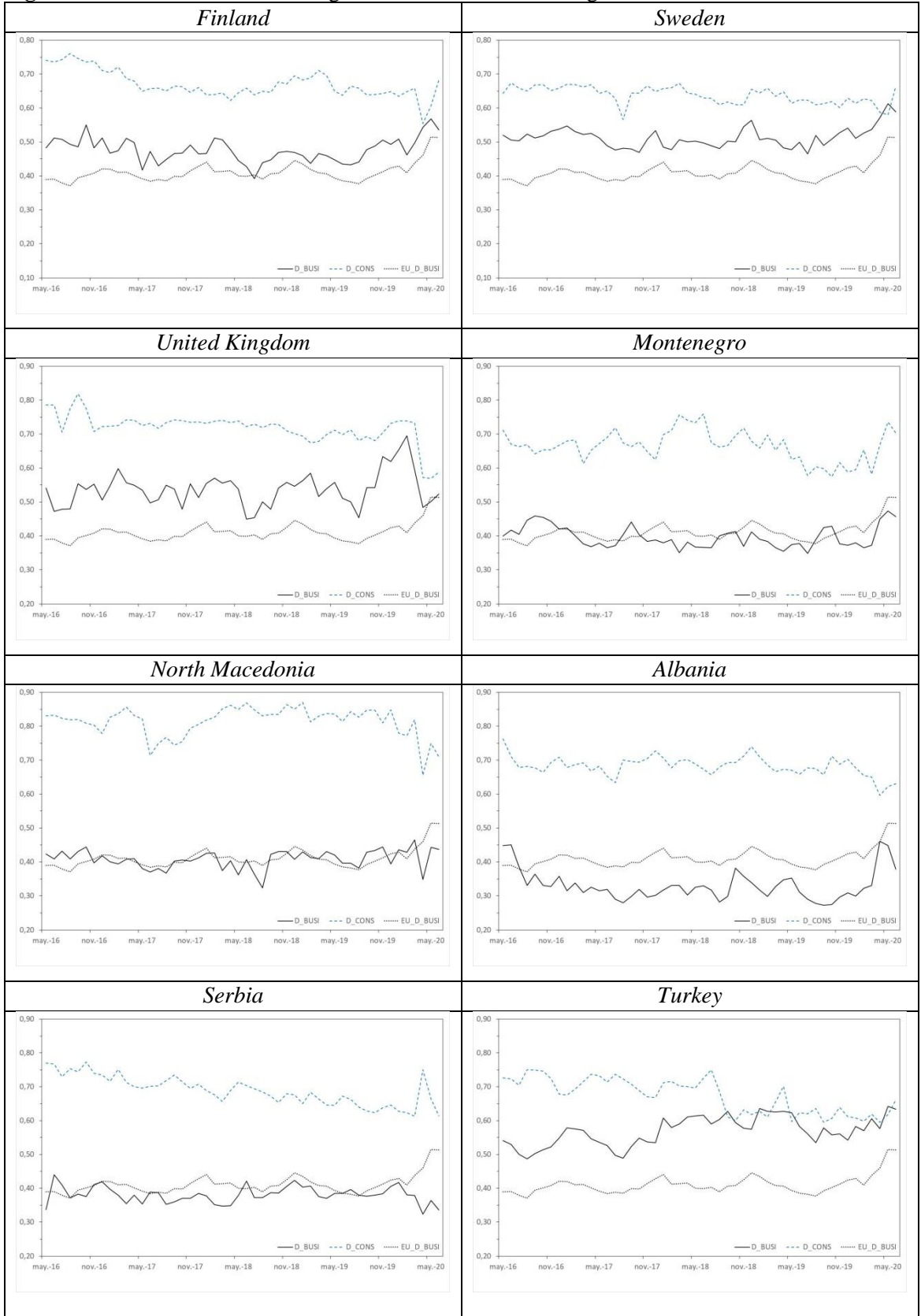

Notes: The solid black line represents the evolution of business disagreement in each country –aggregate disagreement for industry, service, retail trade and construction–, the dashed blue line the evolution of consumer disagreement in each country, and the dotted black line the evolution of aggregate business disagreement in the EU.

**Fig. 2e.** Evolution of business disagreement vs. consumer disagreement

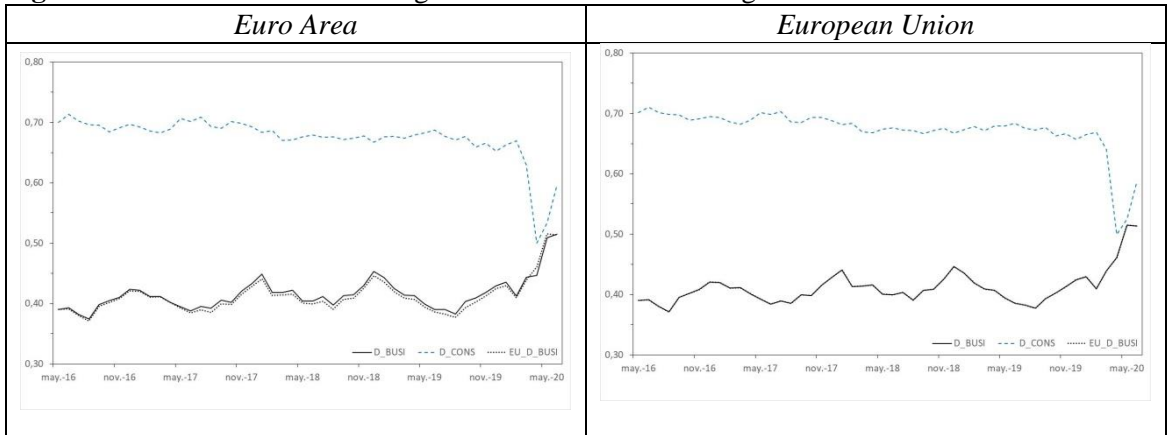

Notes: The solid black line represents the evolution of business disagreement in each country –aggregate disagreement for industry, service, retail trade and construction–, the dashed blue line the evolution of consumer disagreement in each country, and the dotted black line the evolution of aggregate business disagreement in the EU.

In Fig.2 we observe:

- Acute differences across countries in the interdependencies in time of business and consumer disagreement.

- In most countries, the correlation between business and consumer disagreement changes sign during the peak of the pandemic.

- In most countries, around February 2020 consumer disagreement started to decrease sharply, while business disagreement had been rising since mid and late 2019. In Denmark, the Netherlands and Sweden these opposite trends in disagreement are mild.

- In Germany, Croatia, Italy, the UK, Montenegro, North Macedonia and Turkey, the evolution between both types of disagreement during the first half of 2020 seems to positively correlate.

We computed the average values of disagreement for each sector and ranked the countries in increasing order based on the obtained values. Results are presented in Table 2, where we observe that:

- The highest average values of disagreement in the industry and the retail trade sectors are obtained in the United Kingdom, and in the service and the building sectors in Turkey. In Sweden and Ireland we found high levels of disagreement.

- On the other extreme, Portugal, Romania and Poland are the countries in which we obtained the lowest average values of disagreement. An exception is consumer disagreement, where the lowest values where found for Germany.

**Table 2.** Ranking of countries according to the average degree of disagreement by sector

| D_INDU | D_SERV | D_RETA | D_BUIL | D_BUSI | D_CONS | D_TOTAL |
|---|---|---|---|---|---|---|
| Portugal | Romania | Portugal | Italy | Portugal | Germany | Portugal |
| Romania | Poland | Denmark | Cyprus | Romania | Hungary | Poland |
| Cyprus | Latvia | Austria | Belgium | Poland | Poland | Italy |
| Bulgaria | Bulgaria | Lithuania | Portugal | Cyprus | Sweden | Bulgaria |
| Italy | Italy | Poland | Albania | Bulgaria | Italy | Romania |
| Poland | Denmark | Belgium | Serbia | Italy | Bulgaria | Hungary |
| Montenegro | Albania | Malta | Bulgaria | Albania | Lithuania | Denmark |
| Latvia | Cyprus | Czech Rep. | Poland | Belgium | Austria | Cyprus |
| Spain | Portugal | Bulgaria | Romania | Denmark | Portugal | Germany |
| Czech Rep. | Czech Rep. | Cyprus | Czech Rep. | Czech Rep. | **EU** | Albania |
| Albania | Hungary | Netherlands | Denmark | Latvia | **Euro Area** | Latvia |
| Netherlands | Belgium | Albania | Macedonia | Serbia | Romania | Czech Rep. |
| Belgium | Slovenia | Romania | Slovenia | Austria | Latvia | Lithuania |
| Serbia | Macedonia | Latvia | Montenegro | Slovenia | Denmark | Austria |
| Greece | Serbia | Hungary | Germany | Netherlands | France | Netherlands |
| Germany | Lithuania | France | Spain | Montenegro | Turkey | Malta |
| **EU** | Netherlands | Slovenia | Malta | Lithuania | Czech Rep. | Belgium |
| **Euro Area** | **EU** | Croatia | Croatia | Malta | Montenegro | Montenegro |
| Austria | France | Italy | **EU** | Hungary | Estonia | Serbia |
| Macedonia | **Euro Area** | **EU** | Austria | Macedonia | Finland | Greece |
| Lithuania | Croatia | **Euro Area** | **Euro Area** | Spain | Albania | **EU** |
| Hungary | Finland | Serbia | Slovakia | **EU** | Croatia | Slovenia |
| Slovenia | Montenegro | Germany | Hungary | **Euro Area** | Serbia | **Euro Area** |
| Denmark | Greece | Montenegro | Netherlands | Germany | Slovakia | Estonia |
| Malta | Austria | Ireland | France | Croatia | Slovenia | France |
| Croatia | Spain | Slovakia | Lithuania | France | Netherlands | Slovakia |
| Estonia | Estonia | Spain | UK | Slovakia | Belgium | Croatia |
| Slovakia | Malta | Macedonia | Sweden | Greece | Cyprus | Spain |
| Ireland | Germany | Greece | Estonia | Finland | Malta | Finland |
| Turkey | Ireland | Finland | Latvia | Estonia | Spain | Sweden |
| Finland | Slovakia | Sweden | Finland | Ireland | Greece | Macedonia |
| Sweden | UK | Estonia | Ireland | Sweden | UK | Turkey |
| France | Sweden | Turkey | Greece | UK | Ireland | Ireland |
| UK | Turkey | UK | Turkey | Turkey | Macedonia | UK |

Notes: D_INDU refers to the aggregate indicator for the industry, D_SERV for the service sector, D_RETA for the retail trade sector, D_BUIL for the building sector; D_BUSI refers to the business disagreement indicator and is obtained as the arithmetic mean of all aggregate sector indicators, and D_TOTAL averages the business and the consumer discrepancy indicators. Data for the building survey for the UK finishes in November 2019.

Finally, in Fig.3 we graphed for each survey, the evolution of disagreement across the different questions in each sector. We focused the analysis to the EU.

**Fig. 3.** Evolution of disagreement across questions and surveys in the EU

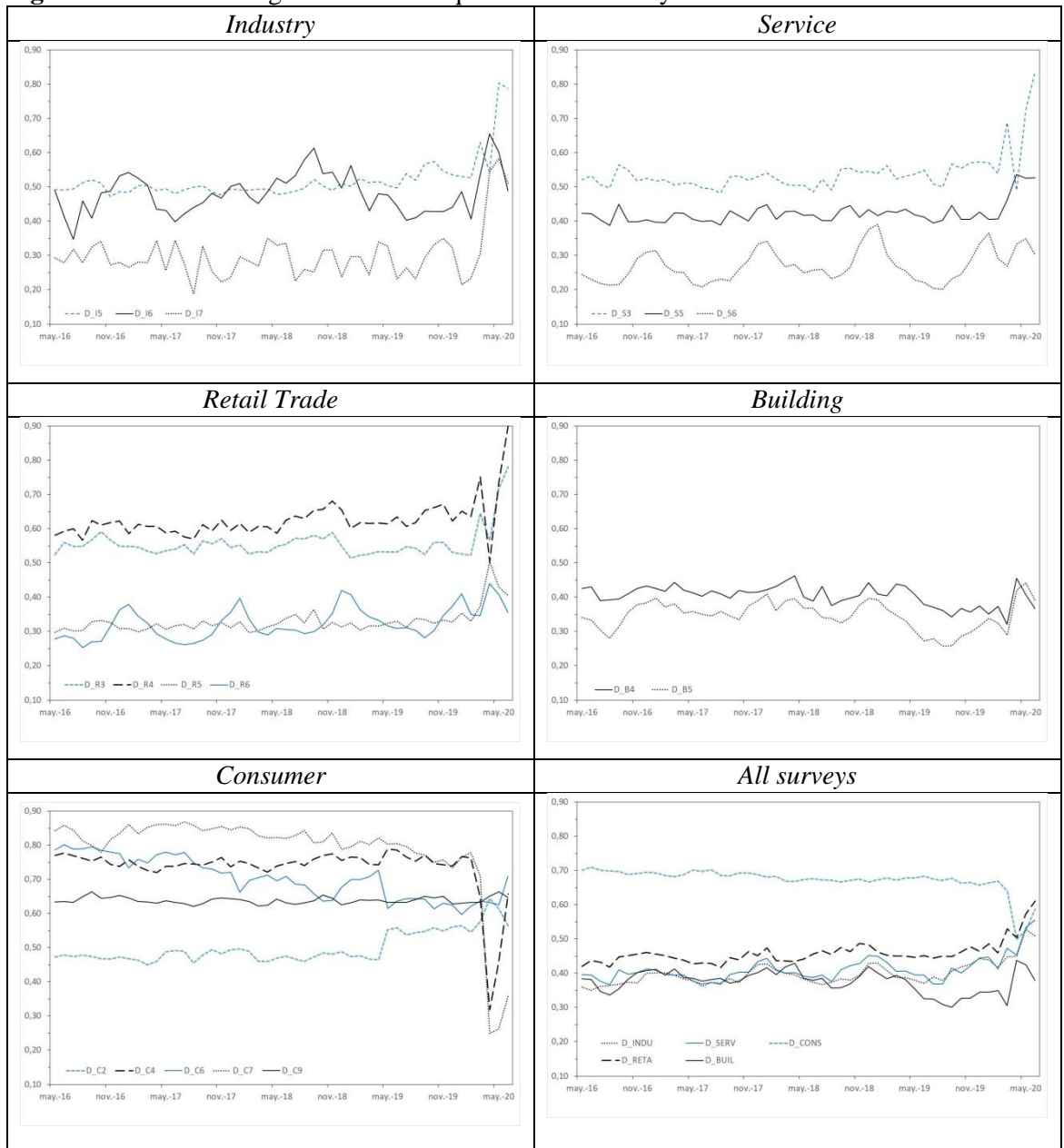

Notes: Each line represents the evolution of disagreement for a specific survey indicator as noted in the legend.

We observe that in most sectors, the evolution of disagreement across sectors follows a similar pattern since the start of the pandemic. The two main exceptions are consumers and the retail trade sector, where we observe negative correlations in the evolution of disagreement across questions. Specifically, we observe a great jump in the level of disagreement in the expectations of business activity in the retail trade sector (D_*R4*) in April 2020. This evolution is similar to the disagreement in the expectations of orders in the retail trade sector (D_*R3*), and opposed to that of prices and employment in the sector.

Similarly, in Fig.3 we also observe that consumers' perception of uncertainty, as captured by the metric of disagreement, strongly diverges across questions. While disagreement about the expected general economic situation (D_*C4*) and the expected unemployment (D_*C7*) experienced a major downturn at the beginning of the pandemic and then a sudden recovery in April 2020, disagreement regarding expectations about price trends (D_*C6*) and major purchases (D_*C9*) experienced the opposite evolution.

When we compare the evolution of aggregate disagreement indicators for the respective sectors and consumers, we find inverse trajectories between firms and consumers. Claveria (2020) obtained similar results for the manufacturing firms and consumers. However, since May 2020 we observe a divergent evolution in industry and construction, which begins to decrease, while uncertainty in service and retail continues to increase.

The obtained results provide a snapshot of economic uncertainty –proxied via indicators of disagreement in business and consumer surveys– in European countries in the midst of the pandemic. These findings give insight regarding the different evolution of uncertainty across economic sectors, variables, agents and countries.

# Appendix

Tables A1 to A4 contain the main descriptive statistics for all variables and surveys.

**Table A1.a.** Descriptive analysis – Average disagreement

|  | D_I5 | D_I6 | D_I7 | D_S3 | D_S5 | D_S6 |
|---|---|---|---|---|---|---|
| Belgium | 0,455 | 0,306 | 0,301 | 0,502 | 0,436 | 0,197 |
| Bulgaria | 0,445 | 0,135 | 0,316 | 0,432 | 0,306 | 0,128 |
| Czech Rep. | 0,403 | 0,267 | 0,372 | 0,434 | 0,451 | 0,166 |
| Denmark | 0,547 | 0,269 | 0,464 | 0,379 | 0,356 | 0,153 |
| Germany | 0,501 | 0,320 | 0,372 | 0,555 | 0,423 | 0,378 |
| Estonia | 0,655 | 0,315 | 0,382 | 0,616 | 0,379 | 0,339 |
| Ireland | 0,571 | 0,369 | 0,469 | 0,596 | 0,477 | 0,352 |
| Greece | 0,604 | 0,240 | 0,346 | 0,552 | 0,445 | 0,315 |
| Spain | 0,433 | 0,267 | 0,324 | 0,639 | 0,414 | 0,272 |
| France | 0,640 | 0,371 | 0,492 | 0,551 | 0,450 | 0,241 |
| Croatia | 0,618 | 0,212 | 0,519 | 0,615 | 0,418 | 0,225 |
| Italy | 0,427 | 0,193 | 0,286 | 0,411 | 0,323 | 0,151 |
| Cyprus | 0,611 | 0,135 | 0,088 | 0,592 | 0,246 | 0,166 |
| Latvia | 0,496 | 0,239 | 0,270 | 0,392 | 0,275 | 0,148 |
| Lithuania | 0,629 | 0,296 | 0,298 | 0,541 | 0,489 | 0,189 |
| Hungary | 0,549 | 0,298 | 0,410 | 0,513 | 0,301 | 0,285 |
| Malta | 0,558 | 0,328 | 0,412 | 0,574 | 0,476 | 0,297 |
| Netherlands | 0,470 | 0,232 | 0,352 | 0,558 | 0,441 | 0,226 |
| Austria | 0,487 | 0,327 | 0,400 | 0,475 | 0,508 | 0,333 |
| Poland | 0,464 | 0,233 | 0,275 | 0,332 | 0,272 | 0,172 |
| Portugal | 0,302 | 0,185 | 0,191 | 0,510 | 0,369 | 0,161 |
| Romania | 0,368 | 0,176 | 0,218 | 0,343 | 0,212 | 0,153 |
| Slovenia | 0,573 | 0,246 | 0,438 | 0,506 | 0,458 | 0,185 |
| Slovakia | 0,688 | 0,227 | 0,469 | 0,640 | 0,567 | 0,238 |
| Finland | 0,570 | 0,390 | 0,472 | 0,540 | 0,465 | 0,263 |
| Sweden | 0,580 | 0,387 | 0,527 | 0,575 | 0,669 | 0,359 |
| United Kingdom | 0,656 | 0,442 | 0,497 | 0,623 | 0,538 | 0,403 |
| Montenegro | 0,666 | 0,574 | 0,156 | 0,681 | 0,393 | 0,198 |
| North Macedonia | 0,621 | 0,209 | 0,387 | 0,661 | 0,307 | 0,186 |
| Albania | 0,593 | 0,145 | 0,303 | 0,568 | 0,249 | 0,089 |
| Serbia | 0,630 | 0,150 | 0,338 | 0,659 | 0,384 | 0,168 |
| Turkey | 0,629 | 0,482 | 0,314 | 0,704 | 0,513 | 0,482 |
| Euro Area | 0,524 | 0,301 | 0,381 | 0,553 | 0,425 | 0,277 |
| European Union | 0,520 | 0,296 | 0,380 | 0,424 | 0,271 | 0,412 |

Notes: D_ denotes discrepancy. D_I5 refers to disagreement in industry production expectations, D_I6 in industry selling price expectations, and D_I7 in industry employment; D_S3 refers to disagreement in service demand expectations, D_S5 in service employment expectations, and D_S6 in service prices expectations.

**Table A1.b.** Descriptive analysis – Average disagreement

|  | D_R3 | D_R4 | D_R5 | D_R6 | D_B4 | D_B5 |
|---|---|---|---|---|---|---|
| Belgium | 0,368 | 0,471 | 0,244 | 0,338 | 0,249 | 0,235 |
| Bulgaria | 0,490 | 0,600 | 0,255 | 0,176 | 0,404 | 0,164 |
| Czech Rep. | 0,366 | 0,468 | 0,383 | 0,259 | 0,359 | 0,275 |
| Denmark | 0,369 | 0,506 | 0,227 | 0,146 | 0,387 | 0,250 |
| Germany | 0,579 | 0,567 | 0,326 | 0,463 | 0,307 | 0,386 |
| Estonia | 0,744 | 0,756 | 0,331 | 0,498 | 0,479 | 0,462 |
| Ireland | 0,682 | 0,653 | 0,321 | 0,349 | 0,552 | 0,494 |
| Greece | 0,758 | 0,768 | 0,322 | 0,233 | 0,595 | 0,475 |
| Spain | 0,688 | 0,702 | 0,323 | 0,328 | 0,467 | 0,233 |
| France | 0,498 | 0,579 | 0,351 | 0,250 | 0,474 | 0,392 |
| Croatia | 0,631 | 0,615 | 0,335 | 0,228 | 0,447 | 0,273 |
| Italy | 0,578 | 0,640 | 0,368 | 0,226 | 0,298 | 0,143 |
| Cyprus | 0,625 | 0,667 | 0,034 | 0,243 | 0,198 | 0,250 |
| Latvia | 0,497 | 0,560 | 0,288 | 0,312 | 0,574 | 0,374 |
| Lithuania | 0,474 | 0,372 | 0,358 | 0,155 | 0,533 | 0,350 |
| Hungary | 0,524 | 0,547 | 0,209 | 0,393 | 0,423 | 0,424 |
| Malta | 0,407 | 0,509 | 0,272 | 0,245 | 0,393 | 0,317 |
| Netherlands | 0,497 | 0,627 | 0,291 | 0,181 | 0,407 | 0,448 |
| Austria | 0,363 | 0,385 | 0,209 | 0,345 | 0,437 | 0,316 |
| Poland | 0,408 | 0,463 | 0,227 | 0,296 | 0,284 | 0,288 |
| Portugal | 0,244 | 0,248 | 0,142 | 0,113 | 0,323 | 0,188 |
| Romania | 0,490 | 0,517 | 0,289 | 0,330 | 0,326 | 0,280 |
| Slovenia | 0,459 | 0,595 | 0,439 | 0,263 | 0,465 | 0,215 |
| Slovakia | 0,618 | 0,605 | 0,454 | 0,327 | 0,429 | 0,384 |
| Finland | 0,551 | 0,699 | 0,407 | 0,471 | 0,549 | 0,401 |
| Sweden | 0,687 | 0,617 | 0,463 | 0,441 | 0,512 | 0,428 |
| United Kingdom | 0,681 | 0,697 | 0,558 | 0,572 | 0,419 | 0,478 |
| Montenegro | 0,705 | 0,682 | 0,330 | 0,251 | 0,484 | 0,200 |
| North Macedonia | 0,753 | 0,752 | 0,240 | 0,310 | 0,385 | 0,277 |
| Albania | 0,606 | 0,655 | 0,138 | 0,197 | 0,393 | 0,161 |
| Serbia | 0,681 | 0,688 | 0,275 | 0,224 | 0,440 | 0,124 |
| Turkey | 0,692 | 0,751 | 0,363 | 0,574 | 0,677 | 0,608 |
| Euro Area | 0,566 | 0,631 | 0,333 | 0,325 | 0,406 | 0,350 |
| European Union | 0,558 | 0,625 | 0,329 | 0,325 | 0,405 | 0,348 |

Notes: D_ denotes discrepancy. D_R3 refers to disagreement in retail trade order expectations, D_R4 in retail trade business activity expectations, D_R5 in retail trade employment expectations, and D_R6 in retail trade prices expectations; D_B4 refers to disagreement in in construction employment expectations, and D_B5 in construction prices expectations.

**Table A1.c.** Descriptive analysis – Average disagreement

|  | D_C2 | D_C4 | D_C6 | D_C7 | D_C9 | D_CONS |
|---|---|---|---|---|---|---|
| Belgium | 0,438 | 0,813 | 0,694 | 0,836 | 0,792 | 0,715 |
| Bulgaria | 0,641 | 0,658 | 0,585 | 0,625 | 0,697 | 0,641 |
| Czech Rep. | 0,615 | 0,781 | 0,476 | 0,722 | 0,852 | 0,689 |
| Denmark | 0,651 | 0,762 | 0,588 | 0,745 | 0,544 | 0,658 |
| Germany | 0,372 | 0,579 | 0,678 | 0,688 | 0,630 | 0,589 |
| Estonia | 0,557 | 0,714 | 0,508 | 0,644 | 0,619 | 0,609 |
| Ireland | 0,644 | 0,730 | 0,756 | 0,793 | 0,849 | 0,754 |
| Greece | 0,543 | 0,538 | 0,778 | 0,606 | 0,537 | 0,600 |
| Spain | 0,587 | 0,815 | 0,792 | 0,813 | 0,630 | 0,727 |
| France | 0,576 | 0,743 | 0,610 | 0,795 | 0,561 | 0,657 |
| Croatia | 0,579 | 0,738 | 0,792 | 0,865 | 0,572 | 0,709 |
| Italy | 0,329 | 0,720 | 0,613 | 0,719 | 0,562 | 0,589 |
| Cyprus | 0,506 | 0,683 | 0,803 | 0,762 | 0,725 | 0,696 |
| Latvia | 0,559 | 0,687 | 0,735 | 0,645 | 0,668 | 0,659 |
| Lithuania | 0,541 | 0,735 | 0,515 | 0,748 | 0,656 | 0,639 |
| Hungary | 0,535 | 0,616 | 0,572 | 0,610 | 0,480 | 0,563 |
| Malta | 0,528 | 0,589 | 0,858 | 0,590 | 0,667 | 0,646 |
| Netherlands | 0,657 | 0,706 | 0,512 | 0,685 | 0,676 | 0,647 |
| Austria | 0,541 | 0,757 | 0,478 | 0,760 | 0,746 | 0,657 |
| Poland | 0,483 | 0,666 | 0,762 | 0,529 | 0,507 | 0,589 |
| Portugal | 0,461 | 0,700 | 0,663 | 0,758 | 0,614 | 0,639 |
| Romania | 0,622 | 0,693 | 0,672 | 0,621 | 0,696 | 0,661 |
| Slovenia | 0,645 | 0,791 | 0,505 | 0,832 | 0,702 | 0,695 |
| Slovakia | 0,500 | 0,638 | 0,628 | 0,733 | 0,684 | 0,636 |
| Finland | 0,588 | 0,687 | 0,593 | 0,756 | 0,729 | 0,671 |
| Sweden | 0,500 | 0,616 | 0,665 | 0,764 | 0,645 | 0,638 |
| United Kingdom | 0,860 | 0,673 | 0,551 | 0,733 | 0,774 | 0,718 |
| Montenegro | 0,552 | 0,684 | 0,723 | 0,774 | 0,595 | 0,667 |
| North Macedonia | 0,796 | 0,833 | 0,801 | 0,898 | 0,739 | 0,814 |
| Albania | 0,619 | 0,744 | 0,698 | 0,823 | 0,529 | 0,683 |
| Serbia | 0,731 | 0,637 | 0,751 | 0,735 | 0,592 | 0,689 |
| Turkey | 0,740 | 0,828 | 0,783 | 0,572 | 0,443 | 0,673 |
| Euro Area | 0,488 | 0,740 | 0,703 | 0,799 | 0,639 | 0,674 |
| European Union | 0,500 | 0,734 | 0,704 | 0,786 | 0,639 | 0,673 |

Notes: D_ denotes discrepancy. D_$C2$ refers to disagreement in consumers' financial situation expectations, D_$C4$ in general economic situation expectations, D_$C6$ in price trends expectations; D_$C7$ in unemployment expectations, and D_$C9$ in consumers' major purchases expectations; D_CONS is computed as the arithmetic mean of the five indicators for consumers.

**Table A2.a.** Descriptive analysis – Minimum disagreement

|  | D_I5 | D_I6 | D_I7 | D_S3 | D_S5 | D_S6 |
|---|---|---|---|---|---|---|
| Belgium | 0,331 | 0,163 | 0,187 | 0,318 | 0,294 | 0,055 |
| Bulgaria | 0,380 | 0,072 | 0,190 | 0,316 | 0,178 | 0,057 |
| Czech Rep. | 0,233 | 0,144 | 0,277 | 0,339 | 0,209 | 0,060 |
| Denmark | 0,351 | 0,145 | 0,250 | 0,290 | 0,242 | 0,093 |
| Germany | 0,363 | 0,181 | 0,270 | 0,426 | 0,349 | 0,296 |
| Estonia | 0,508 | 0,160 | 0,255 | 0,274 | 0,274 | 0,205 |
| Ireland | 0,437 | 0,161 | 0,169 | 0,408 | 0,320 | 0,219 |
| Greece | 0,469 | 0,139 | 0,161 | 0,382 | 0,274 | 0,116 |
| Spain | 0,322 | 0,143 | 0,225 | 0,442 | 0,287 | 0,187 |
| France | 0,389 | 0,237 | 0,429 | 0,446 | 0,345 | 0,166 |
| Croatia | 0,522 | 0,101 | 0,344 | 0,321 | 0,277 | 0,132 |
| Italy | 0,372 | 0,134 | 0,237 | 0,288 | 0,184 | 0,094 |
| Cyprus | 0,218 | 0,032 | 0,000 | 0,099 | 0,042 | 0,074 |
| Latvia | 0,434 | 0,171 | 0,204 | 0,286 | 0,182 | 0,087 |
| Lithuania | 0,487 | 0,204 | 0,187 | 0,378 | 0,353 | 0,115 |
| Hungary | 0,424 | 0,150 | 0,243 | 0,138 | 0,096 | 0,064 |
| Malta | 0,188 | 0,069 | 0,132 | 0,413 | 0,254 | 0,108 |
| Netherlands | 0,388 | 0,123 | 0,259 | 0,414 | 0,367 | 0,108 |
| Austria | 0,389 | 0,153 | 0,308 | 0,334 | 0,417 | 0,181 |
| Poland | 0,396 | 0,145 | 0,207 | 0,278 | 0,208 | 0,109 |
| Portugal | 0,235 | 0,115 | 0,151 | 0,395 | 0,299 | 0,087 |
| Romania | 0,281 | 0,061 | 0,150 | 0,241 | 0,137 | 0,065 |
| Slovenia | 0,457 | 0,170 | 0,341 | 0,383 | 0,348 | 0,120 |
| Slovakia | 0,362 | 0,075 | 0,285 | 0,300 | 0,352 | 0,060 |
| Finland | 0,428 | 0,269 | 0,360 | 0,440 | 0,312 | 0,120 |
| Sweden | 0,469 | 0,209 | 0,374 | 0,411 | 0,524 | 0,228 |
| United Kingdom | 0,351 | 0,215 | 0,368 | 0,149 | 0,345 | 0,256 |
| Montenegro | 0,420 | 0,034 | 0,129 | 0,414 | 0,248 | 0,078 |
| North Macedonia | 0,355 | 0,123 | 0,210 | 0,575 | 0,122 | 0,079 |
| Albania | 0,597 | 0,470 | 0,060 | 0,436 | 0,100 | 0,016 |
| Serbia | 0,521 | 0,087 | 0,192 | 0,543 | 0,191 | 0,088 |
| Turkey | 0,492 | 0,347 | 0,236 | 0,435 | 0,414 | 0,328 |
| Euro Area | 0,467 | 0,222 | 0,334 | 0,488 | 0,383 | 0,207 |
| European Union | 0,472 | 0,220 | 0,337 | 0,482 | 0,387 | 0,201 |

Notes: D_ denotes discrepancy. D_$I5$ refers to disagreement in industry production expectations, D_$I6$ in industry selling price expectations, and D_$I7$ in industry employment; D_$S3$ refers to disagreement in service demand expectations, D_$S5$ in service employment expectations, and D_$S6$ in service prices expectations.

**Table A2.b.** Descriptive analysis – Minimum disagreement

|  | D_R3 | D_R4 | D_R5 | D_R6 | D_B4 | D_B5 |
|---|---|---|---|---|---|---|
| Belgium | 0,225 | 0,297 | 0,133 | 0,149 | 0,177 | 0,154 |
| Bulgaria | 0,294 | 0,465 | 0,124 | 0,100 | 0,276 | 0,094 |
| Czech Rep. | 0,196 | 0,298 | 0,269 | 0,097 | 0,191 | 0,147 |
| Denmark | 0,222 | 0,349 | 0,126 | 0,093 | 0,296 | 0,169 |
| Germany | 0,492 | 0,418 | 0,264 | 0,342 | 0,209 | 0,252 |
| Estonia | 0,564 | 0,499 | 0,179 | 0,225 | 0,358 | 0,296 |
| Ireland | 0,490 | 0,430 | 0,159 | 0,190 | 0,398 | 0,417 |
| Greece | 0,540 | 0,548 | 0,065 | 0,029 | 0,160 | 0,216 |
| Spain | 0,477 | 0,459 | 0,243 | 0,211 | 0,124 | 0,052 |
| France | 0,385 | 0,447 | 0,264 | 0,166 | 0,368 | 0,282 |
| Croatia | 0,392 | 0,392 | 0,210 | 0,119 | 0,337 | 0,105 |
| Italy | 0,401 | 0,498 | 0,256 | 0,165 | 0,229 | 0,081 |
| Cyprus | 0,417 | 0,427 | 0,006 | 0,058 | 0,037 | 0,105 |
| Latvia | 0,207 | 0,451 | 0,199 | 0,222 | 0,443 | 0,282 |
| Lithuania | 0,318 | 0,256 | 0,208 | 0,080 | 0,353 | 0,222 |
| Hungary | 0,054 | 0,049 | 0,069 | 0,107 | 0,305 | 0,239 |
| Malta | 0,083 | 0,111 | 0,038 | 0,016 | 0,133 | 0,097 |
| Netherlands | 0,355 | 0,495 | 0,195 | 0,096 | 0,255 | 0,195 |
| Austria | 0,239 | 0,210 | 0,079 | 0,176 | 0,249 | 0,157 |
| Poland | 0,320 | 0,376 | 0,171 | 0,180 | 0,196 | 0,177 |
| Portugal | 0,168 | 0,183 | 0,093 | 0,073 | 0,181 | 0,120 |
| Romania | 0,250 | 0,325 | 0,141 | 0,068 | 0,207 | 0,124 |
| Slovenia | 0,264 | 0,441 | 0,188 | 0,079 | 0,283 | 0,080 |
| Slovakia | 0,434 | 0,424 | 0,311 | 0,135 | 0,300 | 0,089 |
| Finland | 0,368 | 0,405 | 0,180 | 0,308 | 0,395 | 0,180 |
| Sweden | 0,523 | 0,498 | 0,329 | 0,235 | 0,311 | 0,279 |
| United Kingdom | 0,273 | 0,277 | 0,196 | 0,373 | 0,186 | 0,184 |
| Montenegro | 0,555 | 0,433 | 0,225 | 0,149 | 0,320 | 0,086 |
| North Macedonia | 0,145 | 0,203 | 0,396 | 0,271 | 0,228 | 0,324 |
| Albania | 0,477 | 0,456 | 0,061 | 0,111 | 0,254 | 0,060 |
| Serbia | 0,544 | 0,466 | 0,156 | 0,127 | 0,237 | 0,049 |
| Turkey | 0,428 | 0,420 | 0,274 | 0,431 | 0,566 | 0,432 |
| Euro Area | 0,519 | 0,494 | 0,292 | 0,260 | 0,316 | 0,249 |
| European Union | 0,514 | 0,504 | 0,297 | 0,254 | 0,322 | 0,258 |

Notes: D_ denotes discrepancy. D_R3 refers to disagreement in retail trade order expectations, D_R4 in retail trade business activity expectations, D_R5 in retail trade employment expectations, and D_R6 in retail trade prices expectations; D_B4 refers to disagreement in in construction employment expectations, and D_B5 in construction prices expectations.

**Table A2.c.** Descriptive analysis – Minimum disagreement

|  | D_C2 | D_C4 | D_C6 | D_C7 | D_C9 | D_CONS |
|---|---|---|---|---|---|---|
| Belgium | 0,344 | 0,410 | 0,532 | 0,189 | 0,693 | 0,506 |
| Bulgaria | 0,563 | 0,337 | 0,455 | 0,177 | 0,541 | 0,464 |
| Czech Rep. | 0,532 | 0,447 | 0,349 | 0,285 | 0,761 | 0,553 |
| Denmark | 0,597 | 0,637 | 0,463 | 0,421 | 0,459 | 0,597 |
| Germany | 0,236 | 0,367 | 0,441 | 0,221 | 0,588 | 0,485 |
| Estonia | 0,361 | 0,306 | 0,247 | 0,250 | 0,360 | 0,458 |
| Ireland | 0,591 | 0,283 | 0,614 | 0,382 | 0,628 | 0,559 |
| Greece | 0,328 | 0,227 | 0,559 | 0,187 | 0,371 | 0,384 |
| Spain | 0,529 | 0,287 | 0,666 | 0,216 | 0,535 | 0,521 |
| France | 0,493 | 0,274 | 0,450 | 0,259 | 0,488 | 0,468 |
| Croatia | 0,470 | 0,346 | 0,649 | 0,390 | 0,507 | 0,557 |
| Italy | 0,260 | 0,491 | 0,502 | 0,277 | 0,507 | 0,510 |
| Cyprus | 0,407 | 0,214 | 0,621 | 0,230 | 0,488 | 0,459 |
| Latvia | 0,423 | 0,364 | 0,557 | 0,277 | 0,574 | 0,579 |
| Lithuania | 0,482 | 0,391 | 0,304 | 0,226 | 0,582 | 0,487 |
| Hungary | 0,422 | 0,267 | 0,299 | 0,153 | 0,335 | 0,393 |
| Malta | 0,359 | 0,382 | 0,705 | 0,257 | 0,409 | 0,524 |
| Netherlands | 0,558 | 0,244 | 0,192 | 0,186 | 0,621 | 0,469 |
| Austria | 0,470 | 0,379 | 0,356 | 0,368 | 0,673 | 0,537 |
| Poland | 0,361 | 0,252 | 0,262 | 0,159 | 0,368 | 0,424 |
| Portugal | 0,394 | 0,177 | 0,523 | 0,123 | 0,487 | 0,384 |
| Romania | 0,504 | 0,432 | 0,504 | 0,333 | 0,527 | 0,517 |
| Slovenia | 0,558 | 0,218 | 0,397 | 0,172 | 0,545 | 0,394 |
| Slovakia | 0,437 | 0,266 | 0,424 | 0,168 | 0,616 | 0,409 |
| Finland | 0,538 | 0,565 | 0,417 | 0,269 | 0,637 | 0,555 |
| Sweden | 0,422 | 0,496 | 0,563 | 0,270 | 0,473 | 0,566 |
| United Kingdom | 0,781 | 0,378 | 0,430 | 0,365 | 0,677 | 0,569 |
| Montenegro | 0,380 | 0,537 | 0,577 | 0,555 | 0,519 | 0,575 |
| North Macedonia | 0,608 | 0,580 | 0,661 | 0,641 | 0,475 | 0,656 |
| Albania | 0,543 | 0,678 | 0,561 | 0,519 | 0,409 | 0,597 |
| Serbia | 0,628 | 0,506 | 0,536 | 0,506 | 0,394 | 0,613 |
| Turkey | 0,651 | 0,624 | 0,680 | 0,421 | 0,380 | 0,595 |
| Euro Area | 0,435 | 0,331 | 0,601 | 0,268 | 0,620 | 0,500 |
| European Union | 0,451 | 0,318 | 0,598 | 0,250 | 0,622 | 0,499 |

Notes: D_ denotes discrepancy. D_*C2* refers to disagreement in consumers' financial situation expectations, D_*C4* in general economic situation expectations, D_*C6* in price trends expectations; D_*C7* in unemployment expectations, and D_*C9* in consumers' major purchases expectations; D_CONS is computed as the arithmetic mean of the five indicators for consumers.

**Table A3.a.** Descriptive analysis – Maximum disagreement

|  | D_I5 | D_I6 | D_I7 | D_S3 | D_S5 | D_S6 |
|---|---|---|---|---|---|---|
| Belgium | 0,732 | 0,469 | 0,583 | 0,914 | 0,560 | 0,530 |
| Bulgaria | 0,636 | 0,248 | 0,466 | 0,710 | 0,486 | 0,269 |
| Czech Rep. | 0,778 | 0,365 | 0,542 | 0,820 | 0,628 | 0,457 |
| Denmark | 0,793 | 0,465 | 0,676 | 0,653 | 0,540 | 0,387 |
| Germany | 0,809 | 0,536 | 0,460 | 0,829 | 0,493 | 0,497 |
| Estonia | 0,794 | 0,505 | 0,585 | 0,734 | 0,564 | 0,542 |
| Ireland | 0,848 | 0,610 | 0,682 | 0,882 | 0,696 | 0,519 |
| Greece | 0,950 | 0,383 | 0,636 | 0,816 | 0,654 | 0,548 |
| Spain | 0,767 | 0,449 | 0,542 | 0,855 | 0,579 | 0,455 |
| France | 0,878 | 0,514 | 0,567 | 0,860 | 0,776 | 0,396 |
| Croatia | 0,944 | 0,323 | 0,615 | 0,873 | 0,553 | 0,327 |
| Italy | 0,757 | 0,287 | 0,329 | 0,848 | 0,429 | 0,257 |
| Cyprus | 0,812 | 0,274 | 0,202 | 0,851 | 0,514 | 0,412 |
| Latvia | 0,675 | 0,341 | 0,446 | 0,650 | 0,514 | 0,317 |
| Lithuania | 0,883 | 0,411 | 0,438 | 0,849 | 0,621 | 0,389 |
| Hungary | 0,774 | 0,540 | 0,575 | 0,779 | 0,455 | 0,457 |
| Malta | 0,797 | 0,680 | 0,712 | 0,853 | 0,706 | 0,496 |
| Netherlands | 0,652 | 0,416 | 0,456 | 0,815 | 0,500 | 0,489 |
| Austria | 0,864 | 0,529 | 0,505 | 0,918 | 0,605 | 0,546 |
| Poland | 0,688 | 0,427 | 0,505 | 0,615 | 0,439 | 0,445 |
| Portugal | 0,768 | 0,482 | 0,449 | 0,796 | 0,498 | 0,358 |
| Romania | 0,741 | 0,324 | 0,496 | 0,628 | 0,487 | 0,378 |
| Slovenia | 0,968 | 0,392 | 0,599 | 0,850 | 0,555 | 0,271 |
| Slovakia | 0,980 | 0,531 | 0,636 | 0,885 | 0,878 | 0,512 |
| Finland | 0,766 | 0,525 | 0,652 | 0,713 | 0,662 | 0,522 |
| Sweden | 0,765 | 0,577 | 0,692 | 0,777 | 0,783 | 0,515 |
| United Kingdom | 0,849 | 0,851 | 0,644 | 0,949 | 0,794 | 0,574 |
| Montenegro | 0,959 | 0,285 | 0,466 | 0,968 | 0,574 | 0,401 |
| North Macedonia | 0,811 | 0,328 | 0,488 | 0,834 | 0,413 | 0,257 |
| Albania | 0,718 | 0,276 | 0,482 | 0,876 | 0,425 | 0,218 |
| Serbia | 0,780 | 0,205 | 0,396 | 0,814 | 0,483 | 0,300 |
| Turkey | 0,885 | 0,655 | 0,410 | 0,883 | 0,653 | 0,649 |
| Euro Area | 0,824 | 0,444 | 0,468 | 0,853 | 0,536 | 0,403 |
| European Union | 0,804 | 0,427 | 0,482 | 0,837 | 0,537 | 0,391 |

Notes: D_ denotes discrepancy. D_*I5* refers to disagreement in industry production expectations, D_*I6* in industry selling price expectations, and D_*I7* in industry employment; D_*S3* refers to disagreement in service demand expectations, D_*S5* in service employment expectations, and D_*S6* in service prices expectations.

**Table A3.b.** Descriptive analysis – Maximum disagreement

|  | D_R3 | D_R4 | D_R5 | D_R6 | D_B4 | D_B5 |
|---|---|---|---|---|---|---|
| Belgium | 0,762 | 0,774 | 0,493 | 0,602 | 0,411 | 0,377 |
| Bulgaria | 0,733 | 0,769 | 0,516 | 0,432 | 0,557 | 0,290 |
| Czech Rep. | 0,705 | 0,886 | 0,521 | 0,442 | 0,516 | 0,486 |
| Denmark | 0,610 | 0,700 | 0,520 | 0,314 | 0,564 | 0,495 |
| Germany | 0,753 | 0,785 | 0,493 | 0,606 | 0,420 | 0,522 |
| Estonia | 0,925 | 0,961 | 0,540 | 0,640 | 0,655 | 0,635 |
| Ireland | 0,933 | 0,867 | 0,628 | 0,606 | 0,658 | 0,575 |
| Greece | 0,935 | 0,957 | 0,642 | 0,520 | 0,860 | 0,640 |
| Spain | 0,972 | 0,918 | 0,549 | 0,478 | 0,765 | 0,541 |
| France | 0,667 | 0,888 | 0,480 | 0,430 | 0,563 | 0,558 |
| Croatia | 0,838 | 0,910 | 0,493 | 0,356 | 0,550 | 0,398 |
| Italy | 0,946 | 0,901 | 0,454 | 0,373 | 0,359 | 0,236 |
| Cyprus | 0,801 | 0,827 | 0,130 | 0,475 | 0,412 | 0,452 |
| Latvia | 0,749 | 0,819 | 0,573 | 0,461 | 0,821 | 0,502 |
| Lithuania | 0,759 | 0,666 | 0,537 | 0,446 | 0,704 | 0,478 |
| Hungary | 0,686 | 0,794 | 0,319 | 0,557 | 0,614 | 0,555 |
| Malta | 0,689 | 0,716 | 0,592 | 0,897 | 0,547 | 0,619 |
| Netherlands | 0,687 | 0,786 | 0,476 | 0,484 | 0,538 | 0,534 |
| Austria | 0,584 | 0,691 | 0,463 | 0,591 | 0,608 | 0,492 |
| Poland | 0,643 | 0,706 | 0,480 | 0,623 | 0,506 | 0,581 |
| Portugal | 0,587 | 0,597 | 0,318 | 0,263 | 0,465 | 0,298 |
| Romania | 0,932 | 0,866 | 0,618 | 0,534 | 0,542 | 0,491 |
| Slovenia | 0,791 | 0,915 | 0,755 | 0,611 | 0,578 | 0,352 |
| Slovakia | 0,944 | 0,855 | 0,592 | 0,549 | 0,582 | 0,532 |
| Finland | 0,922 | 0,896 | 0,656 | 0,678 | 0,731 | 0,627 |
| Sweden | 0,902 | 0,865 | 0,713 | 0,808 | 0,630 | 0,583 |
| United Kingdom | 0,897 | 0,930 | 0,786 | 0,770 | 0,554 | 0,629 |
| Montenegro | 0,999 | 0,933 | 0,465 | 0,440 | 0,619 | 0,320 |
| North Macedonia | 0,929 | 0,928 | 0,358 | 0,442 | 0,543 | 0,407 |
| Albania | 0,809 | 0,789 | 0,286 | 0,360 | 0,637 | 0,397 |
| Serbia | 0,863 | 0,894 | 0,423 | 0,360 | 0,545 | 0,250 |
| Turkey | 0,883 | 0,942 | 0,491 | 0,779 | 0,794 | 0,800 |
| Euro Area | 0,802 | 0,911 | 0,504 | 0,424 | 0,470 | 0,426 |
| European Union | 0,782 | 0,902 | 0,504 | 0,441 | 0,463 | 0,442 |

Notes: D_ denotes discrepancy. D_R3 refers to disagreement in retail trade order expectations, D_R4 in retail trade business activity expectations, D_R5 in retail trade employment expectations, and D_R6 in retail trade prices expectations; D_B4 refers to disagreement in in construction employment expectations, and D_B5 in construction prices expectations.

**Table A3.c.** Descriptive analysis – Maximum disagreement

|  | D_C2 | D_C4 | D_C6 | D_C7 | D_C9 | D_CONS |
|---|---|---|---|---|---|---|
| Belgium | 0,536 | 0,965 | 0,788 | 0,981 | 0,854 | 0,791 |
| Bulgaria | 0,920 | 0,733 | 0,802 | 0,745 | 0,761 | 0,700 |
| Czech Rep. | 0,760 | 0,886 | 0,655 | 0,929 | 0,922 | 0,757 |
| Denmark | 0,749 | 0,874 | 0,714 | 0,902 | 0,629 | 0,737 |
| Germany | 0,690 | 0,689 | 0,819 | 0,796 | 0,662 | 0,616 |
| Estonia | 0,675 | 0,919 | 0,980 | 0,743 | 0,773 | 0,761 |
| Ireland | 0,798 | 0,982 | 0,874 | 0,938 | 0,938 | 0,836 |
| Greece | 0,746 | 0,934 | 0,938 | 0,966 | 0,667 | 0,815 |
| Spain | 0,704 | 0,965 | 0,884 | 0,985 | 0,707 | 0,806 |
| France | 0,684 | 0,918 | 0,768 | 0,930 | 0,626 | 0,738 |
| Croatia | 0,737 | 0,811 | 0,958 | 0,980 | 0,660 | 0,770 |
| Italy | 0,525 | 0,965 | 0,763 | 0,898 | 0,610 | 0,679 |
| Cyprus | 0,639 | 0,856 | 0,951 | 0,923 | 0,833 | 0,798 |
| Latvia | 0,715 | 0,809 | 0,973 | 0,759 | 0,817 | 0,722 |
| Lithuania | 0,651 | 0,816 | 0,813 | 0,853 | 0,752 | 0,700 |
| Hungary | 0,831 | 0,760 | 0,806 | 0,705 | 0,680 | 0,625 |
| Malta | 0,679 | 0,911 | 0,996 | 0,871 | 0,976 | 0,805 |
| Netherlands | 0,760 | 0,886 | 0,770 | 0,947 | 0,763 | 0,783 |
| Austria | 0,687 | 0,866 | 0,653 | 0,957 | 0,824 | 0,707 |
| Poland | 0,703 | 0,865 | 0,971 | 0,674 | 0,874 | 0,670 |
| Portugal | 0,610 | 0,840 | 0,714 | 0,935 | 0,662 | 0,708 |
| Romania | 0,763 | 0,781 | 0,953 | 0,683 | 0,783 | 0,719 |
| Slovenia | 0,713 | 0,896 | 0,667 | 0,980 | 0,749 | 0,779 |
| Slovakia | 0,608 | 0,761 | 0,914 | 0,920 | 0,744 | 0,776 |
| Finland | 0,664 | 0,851 | 0,759 | 0,907 | 0,791 | 0,761 |
| Sweden | 0,565 | 0,916 | 0,748 | 0,936 | 0,702 | 0,675 |
| United Kingdom | 0,926 | 0,902 | 0,843 | 0,880 | 0,854 | 0,820 |
| Montenegro | 0,923 | 0,846 | 0,856 | 0,934 | 0,693 | 0,760 |
| North Macedonia | 0,912 | 0,977 | 0,910 | 0,982 | 0,868 | 0,870 |
| Albania | 0,691 | 0,851 | 0,827 | 0,948 | 0,581 | 0,763 |
| Serbia | 0,870 | 0,860 | 0,975 | 0,979 | 0,713 | 0,773 |
| Turkey | 0,810 | 0,989 | 0,923 | 0,745 | 0,607 | 0,751 |
| Euro Area | 0,626 | 0,800 | 0,792 | 0,893 | 0,670 | 0,713 |
| European Union | 0,644 | 0,790 | 0,802 | 0,869 | 0,664 | 0,710 |

Notes: D_ denotes discrepancy. D_*C2* refers to disagreement in consumers' financial situation expectations, D_*C4* in general economic situation expectations, D_*C6* in price trends expectations; D_*C7* in unemployment expectations, and D_*C9* in consumers' major purchases expectations; D_CONS is computed as the arithmetic mean of the five indicators for consumers.

**Table A4.** Descriptive analysis – Average disagreement

|  | D_INDU | D_SERV | D_RETA | D_BUIL | D_BUSI | D_TOTAL |
|---|---|---|---|---|---|---|
| Belgium | 0,354 | 0,379 | 0,355 | 0,242 | 0,333 | 0,524 |
| Bulgaria | 0,298 | 0,289 | 0,380 | 0,284 | 0,313 | 0,477 |
| Czech Rep. | 0,347 | 0,350 | 0,369 | 0,317 | 0,346 | 0,518 |
| Denmark | 0,427 | 0,296 | 0,312 | 0,318 | 0,338 | 0,498 |
| Germany | 0,398 | 0,452 | 0,484 | 0,346 | 0,420 | 0,505 |
| Estonia | 0,451 | 0,445 | 0,582 | 0,471 | 0,487 | 0,548 |
| Ireland | 0,469 | 0,475 | 0,501 | 0,523 | 0,492 | 0,624 |
| Greece | 0,397 | 0,437 | 0,520 | 0,535 | 0,472 | 0,536 |
| Spain | 0,341 | 0,442 | 0,510 | 0,350 | 0,411 | 0,569 |
| France | 0,501 | 0,414 | 0,420 | 0,433 | 0,442 | 0,550 |
| Croatia | 0,450 | 0,420 | 0,452 | 0,360 | 0,420 | 0,565 |
| Italy | 0,302 | 0,295 | 0,453 | 0,221 | 0,317 | 0,453 |
| Cyprus | 0,278 | 0,335 | 0,392 | 0,224 | 0,307 | 0,502 |
| Latvia | 0,335 | 0,272 | 0,414 | 0,474 | 0,374 | 0,516 |
| Lithuania | 0,408 | 0,406 | 0,340 | 0,442 | 0,399 | 0,519 |
| Hungary | 0,419 | 0,367 | 0,418 | 0,424 | 0,407 | 0,485 |
| Malta | 0,433 | 0,449 | 0,358 | 0,355 | 0,399 | 0,523 |
| Netherlands | 0,351 | 0,408 | 0,399 | 0,427 | 0,396 | 0,522 |
| Austria | 0,405 | 0,439 | 0,325 | 0,377 | 0,386 | 0,521 |
| Poland | 0,324 | 0,259 | 0,349 | 0,286 | 0,304 | 0,447 |
| Portugal | 0,226 | 0,347 | 0,187 | 0,256 | 0,254 | 0,446 |
| Romania | 0,254 | 0,236 | 0,407 | 0,303 | 0,300 | 0,477 |
| Slovenia | 0,419 | 0,383 | 0,439 | 0,340 | 0,395 | 0,545 |
| Slovakia | 0,461 | 0,482 | 0,501 | 0,406 | 0,463 | 0,550 |
| Finland | 0,477 | 0,423 | 0,532 | 0,475 | 0,477 | 0,574 |
| Sweden | 0,498 | 0,534 | 0,552 | 0,470 | 0,514 | 0,576 |
| United Kingdom | 0,532 | 0,521 | 0,627 | 0,449 | 0,538 | 0,628 |
| Montenegro | 0,330 | 0,424 | 0,492 | 0,342 | 0,397 | 0,531 |
| North Macedonia | 0,406 | 0,384 | 0,514 | 0,331 | 0,409 | 0,611 |
| Albania | 0,347 | 0,302 | 0,399 | 0,277 | 0,331 | 0,507 |
| Serbia | 0,373 | 0,404 | 0,467 | 0,282 | 0,381 | 0,535 |
| Turkey | 0,475 | 0,566 | 0,595 | 0,643 | 0,570 | 0,621 |
| Euro Area | 0,402 | 0,418 | 0,464 | 0,378 | 0,415 | 0,545 |
| European Union | 0,399 | 0,412 | 0,459 | 0,376 | 0,411 | 0,542 |

Notes: D_INDU refers to the aggregate indicator for the industry, D_SERV for the service sector, D_RETA for the retail trade sector, D_BUIL for the building sector; D_BUSI refers to the business disagreement indicator and is obtained as the arithmetic mean of all aggregate sector indicators, and D_TOTAL averages the business and the consumer discrepancy indicators. Data for the building survey for the UK finishes in November 2019.